\documentclass[letterpaper,twocolumn,10pt]{article}
\usepackage{usenix-2020-09}

\usepackage{graphicx}
\graphicspath{ {./figures/} }

\usepackage{appendix}

\usepackage{pifont}
\newcommand{\cmark}{\ding{51}}%
\newcommand{\xmark}{\ding{55}}%

\begin{document}

\date{}

\title{\Large \bf Synthetic Datasets for Program Similarity Research}

\author{
{\rm Alexander Interrante-Grant}\\
MIT Lincoln Laboratory
\and
{\rm Michael Wang}\\
MIT Lincoln Laboratory
\and
{\rm Lisa Baer}\\
MIT Lincoln Laboratory
\and
{\rm Ryan Whelan}\\
MIT Lincoln Laboratory
\and
{\rm Tim Leek}\\
MIT Lincoln Laboratory
} 

\maketitle

\begin{abstract}
Program similarity has become an increasingly popular area of research with
various security applications such as plagiarism detection, author
identification, and malware analysis \cite{haq2019}. However, program
similarity research faces a few unique dataset quality problems in evaluating
the effectiveness of novel approaches. First, few high-quality datasets for
binary program similarity exist and are widely used in this domain. Second,
there are potentially many different, disparate definitions of what makes one
program ``similar'' to another and in many cases there is often a large
semantic gap between the labels provided by a dataset and any useful notion of
behavioral or semantic similarity.

In this paper, we present HELIX – a framework for generating large, synthetic
program similarity datasets. We also introduce Blind HELIX, a tool built on top
of HELIX for extracting HELIX components from library code automatically using
program slicing. We evaluate HELIX and Blind HELIX by comparing the performance
of program similarity tools on a HELIX dataset to a hand-crafted dataset built
from multiple, disparate notions of program similarity. Using Blind HELIX, we
show that HELIX can generate realistic and useful datasets of virtually
infinite size for program similarity research with ground truth labels that
embody practical notions of program similarity. Finally, we discuss the results
and reason about relative tool ranking.
\end{abstract}

\section{Introduction}

Program similarity is a domain of research that covers a broad range of
applications, including: bug search, malware clustering, malware detection,
malware lineage, patch generation and analysis, porting information across
program versions, and software theft detection \cite{haq2019}. Increasingly,
researchers have relied on machine learning techniques as potential solutions
to problems within the program similarity domain \cite{ucci2019}. With the
explosion in popularity of machine learning applications in this domain,
high-quality datasets are more important than ever to ensure that the community
can reasonably and accurately gauge the performance of one approach versus
another.

Unfortunately, high-quality datasets are difficult to find in the domain of
binary program similarity \cite{smith2020}. Existing datasets often capture
opaque notions of program similarity with questionable relevance to real-world
problems. In the malware analysis domain specifically, prior research has found
a large semantic gap between dataset labels and any useful notion of program
similarity \cite{smith2020}. In other domains with comparable dataset
availability and quality issues, researchers often turn to synthetic data
generation and augmentation strategies with reasonable success \cite{avino2018,
yale2019, dash2019, jaderberg2014, gupta2016, tremblay2018}.

To combat the problem of poor dataset availability and quality, this paper
describes our approach for generating synthetic program similarity datasets by
slicing and recombining existing open-source libraries into samples with known,
configurable ground truth similarity. We evaluate our approach against a
manually-labeled dataset comprised of multiple abstract notions of program
similarity.

The rest of this paper is organized as follows. First, Section
\ref{sec:background} covers background information on existing program
similarity approaches and includes a brief survey of dataset use in related
work. Next, we describe our approach for generating synthetic program
similarity datasets in Section \ref{sec:approach}. We then evaluate our
approach in Section \ref{sec:evaluation} and discuss our results and directions
for future work in Section \ref{sec:discussion}. Finally, we summarize related
work in Section \ref{sec:related-work} and conclude in Section
\ref{sec:conclusion}.

Our primary contributions are:

\begin{itemize}
\item To the best of our knowledge, we are the first to use a program slicing
and re-assembly approach to generate synthetic datasets for program similarity
research. Specifically, we combine slices of open-source library code to
generate samples with ground truth similarity which is based in measurable
program similarity.
\item We present HELIX - an open-source framework for program generation and
mutation geared toward dataset generation for program similarity research that
is generic enough to support any programming language, compiler, and build
system.  HELIX is robust, mature, actively being maintained, and publicly
available at
\url{https://github.com/helix-datasets/helix}.
\item We present Blind HELIX - a tool built on top of HELIX that automatically
extracts functional components from library code using program slicing. Blind
HELIX is available at \url{https://github.com/helix-datasets/blind-helix}.
\item We evaluate our approach using a number of existing program similarity
tools and demonstrate that the datasets generated align well with multiple
abstract notions of program similarity. We also discuss the relative
performance of each of these tools against different notions of program
similarity.
\end{itemize}

\section{Background}
\label{sec:background}

This section covers relevant background information before we describe our
synthetic program similarity dataset generation approach. We start by
discussing prior approaches to program similarity, then cover available
datasets and their use by the community. Finally, we cover some grounding
research in other domains on synthetic data generation.

\subsection{Program Similarity}

Haq et al. conducted a study of program similarity applications, techniques,
and evaluation methodologies. They found that program similarity has been used
for a wide range of applications, such as bug search, malware clustering,
malware detection, malware lineage, patch generation and analysis, porting
information across program versions, and software theft detection
\cite{haq2019}.

One of the simplest methods to determine program similarity is through
byte-wise approximate matching, or so-called fuzzy hashing
\cite{harichandran2016}, of whole programs. However, a survey by Pagani et al.
found that existing fuzzy hashing approaches are limited in the types of
similarity they are able to measure and fairly minor changes in the complex
binary executable format can induce large apparent dissimilarities when using
fuzzy hashing \cite{pagani2018}.

In recent years, many more sophisticated approaches to program similarity have
been applied to the problem of malware analysis involving applications of
machine learning \cite{ucci2019}. The quality and availability of datasets for
program similarity is important for machine learning based approaches in
particular which are unlikely to generalize if poor data are used for training.
In the next section, we discuss existing datasets and cover a number of recent
surveys which have questioned the quality and availability of datasets in this
domain.

\subsection{Dataset Availability and Use}

\subsubsection*{Surveys}

Perhaps the most popular application of program similarity is toward problems
of malware analysis and security. Botacin et al. conducted a survey of malware
research covering nearly 500 papers across major academic security conferences
between the years of 2000 and 2018 \cite{botacin2021}. They summarized what
they observed to be ten major pitfalls and challenges in malware research, four
of which directly relate to dataset use and availability. In particular, they
noted that dataset suitability criteria for particular experiments are
inconsistent, dataset representativity of realistic problems is generally poor,
limited reproducibility stemming from private or underspecified dataset use is
rampant, and dataset labeling, particularly labeling by antivirus labels or
labels derived from antivirus labels, is often opaque and inconsistent.

A similar study focused exclusively on dataset availability and use across a
broad range of digital forensic applications found similar issues in this
domain \cite{grajeda2017}. In particular, of the over 700 papers they reviewed,
they found that only 54.4\% used datasets that already existed at the time of
publication, with the remainder opting to produce their own dataset. They also
mentioned issues with private dataset use and general availability. Another
study of dataset availability in the domain of cybersecurity research found
that only 21.2\% of papers which create their own dataset make that dataset
public \cite{zheng2018}.

It is clear that in order for the fields of computer security and digital
forensics to mature, researchers need more and better standardized options for
datasets across the community. Better dataset availability and more
standardized dataset use would lead to much higher rates of reproducibility in
our research, greater community confidence in novel research, and a much better
understanding of the relative performance of comparable solutions.

\subsubsection*{Datasets}

A list of commonly used datasets created specifically for program similarity
and malware analysis research purposes is included in Table
\ref{tab:existing-datasets}. These datasets were distilled from the above
surveys as well as other recent work in program similarity and malware
analysis.

\begin{table*}[t]
\begin{center}
\begin{tabular}{ c c c c c }
\hline
Name & Year & Type & Labels & Available* \\
\hline
Malware Genome Project \cite{zhou2012} & 2012 & Android Malware & Malware Families & \xmark \\
Drebin \cite{arp2014} & 2014 & Android Malware & Malware Families & \cmark \\
Malicia \cite{nappa2013} & 2013 & Drive-by Download Malware & Malware Families & \xmark \\
Microsoft Kaggle \cite{ronan2018} & 2015 & Malware & Malware Families & \cmark \\
EMBER \cite{anderson2018} & 2017 (2018) & Malware & Benign/Malicious & \cmark \\
T5 \cite{roussev2011} & 2011 & Non-Executable Files & Similarity & \cmark \\
MSX-13 \cite{roussev2013} & 2013 & Microsoft Office Files & Similarity & \cmark \\
\hline
\end{tabular}
\footnotetext[1]{Footnote}
\end{center}
\caption{\label{tab:existing-datasets} Published datasets for program
similarity and malware analysis research. *Publicly available for download
at the time of publication of this work.}
\end{table*}

\subsection{Synthetic Dataset Generation}

Synthetic datasets have become increasingly common across a wide range of
research domains. In applications of deep learning especially, where dataset
sizes must be very large to support training complex models, synthetic dataset
use has become fairly common in domains such as computer vision, environmental
simulation, bioinformatics, natural language processing, and more
\cite{nikolenko2019}.

Numerous novel synthetic dataset generation strategies have been proposed in
recent years across many domains. In research involving healthcare data,
synthetic data generation strategies have been proposed to avoid the many
privacy pitfalls involved with real medical data \cite{avino2018, yale2019,
dash2019}. Synthetic dataset generation is also common in the domain of
computer vision for tasks such as text recognition and general data
augmentation \cite{jaderberg2014, gupta2016, tremblay2018}. In the realm of
security, synthetic dataset generation has even been proposed for evaluating
security software \cite{skopik2014} and for fuzzing \cite{dolan2016}.

Multiple recent studies have evaluated the effectiveness of synthetic data
across many domains and applications and have found it to be an effective tool
for research \cite{hittmeir2019, benaim2020}. Patki et al. even designed and
implemented the first generalized generative model for building synthetic
datasets in arbitrary domains \cite{patki2016}.

\section{Approach}
\label{sec:approach}

In order to generate synthetic datasets for program similarity, we implement a
framework for combining small, labeled pieces of reusable, parameterized
implementations of specific program functionality - referred to here as
components. We then automatically harvest a large corpus of components from
existing open-source libraries using program slicing. Finally, by generating
samples which contain different combinations of components, we are able to
create synthetic datasets of programs which have known similarity based on the
components they share. In the next section, we describe the HELIX framework we
implemented for synthetic dataset generation.

\subsection{The HELIX Framework}

At its core, the HELIX framework defines three major primitives:

\begin{description}
\item[Blueprints] Core project layouts including templated boilerplate and
methods for generating and building artifacts from a set of Components and
Transforms. For example, a C++ project build with CMake.

\item[Components]  Small, configurable pieces of source code that represent a
specific implementation of a specific functionality along with associated
metadata. For example, a specific implementation of downloading a file from a
given URL using the cURL library.

\item[Transforms] Modifications of either source code or a built artifact along
with associated metadata. For example, the GNU Binutil \texttt{strip} which
removes debugging symbols from a compiled binary.
\end{description}

\begin{figure}
\begin{center}
\includegraphics[width=\linewidth]{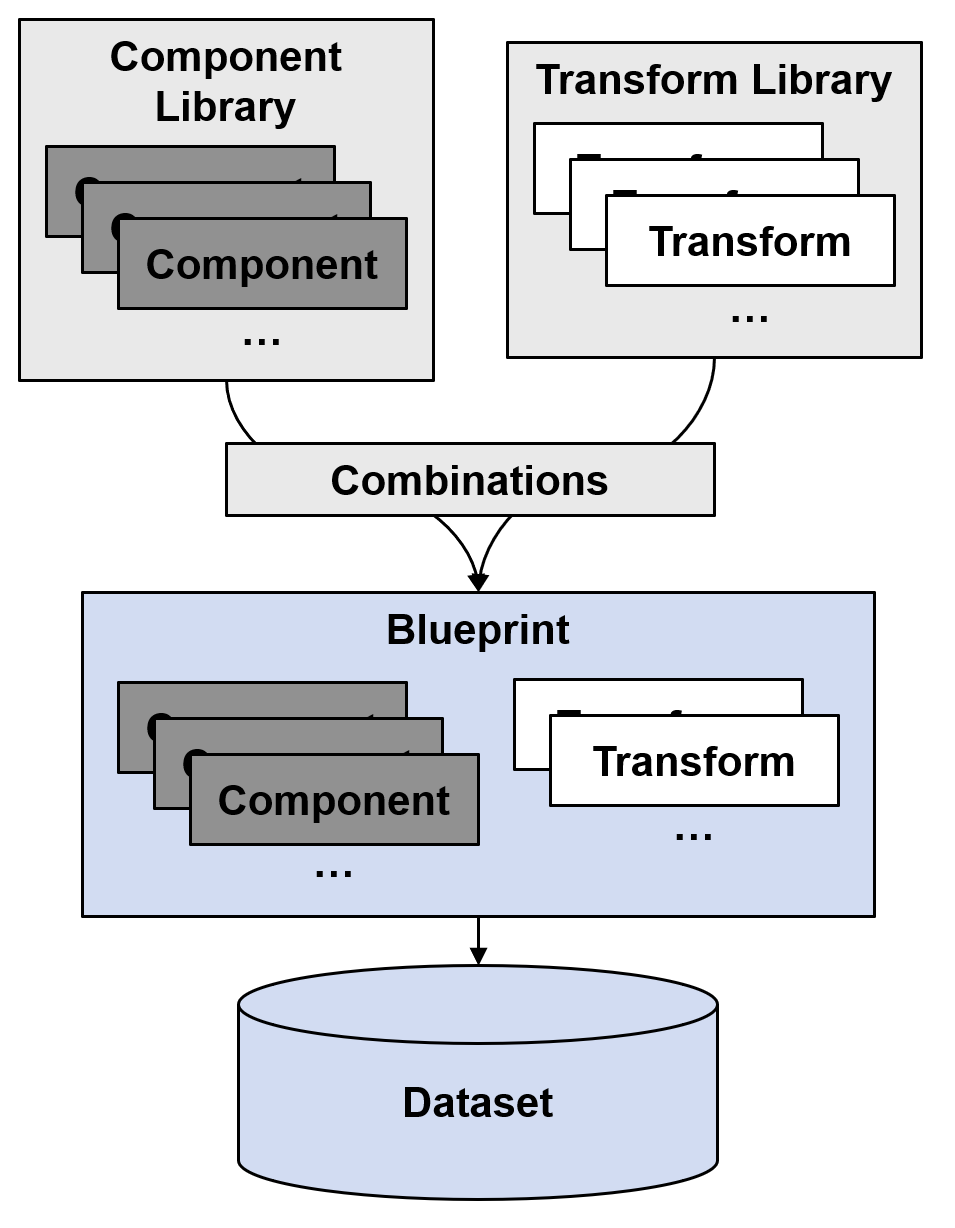}
\end{center}
\caption{\label{fig:helix-design} The design of HELIX - sets of components and
transforms are combined using a blueprint to generate samples in a synthetic
dataset.}
\end{figure}

These core primitives are depicted in Figure \ref{fig:helix-design}. Both
components and transforms are labeled with tags indicating functionality and
all tags associated with included components and transforms are aggregated as
the ground truth labels for a sample in a generated dataset. Pairwise sample
similarity can be computed as a function of the sets of ground truth tags for
each sample. 

For the purposes of this work we make use of components and HELIX's built in
\texttt{CMakeCppBlueprint} to build samples from both C and C++ source code. We
do not make use of transforms but their description is included here for
completeness. The HELIX framework itself is robust and flexible and capable of
generating realistic programs that accomplish meaningful tasks under dynamic
analysis by combining parameterized components in interesting ways. It also
supports a uniform interface for applying transformations and obfuscations to
code and is generic enough to support practically any programming language and
build system.

In the next section, we describe our approach for automatically extracting
HELIX components from existing library source code which allows us to generate
arbitrarily large datasets with HELIX using a tool built on top of HELIX that
we call Blind HELIX.

\subsection{Blind HELIX}
\label{sec:program-slicing-for-components}

While HELIX includes a small number of hand-written components, on their own
they are not enough to generate any sizable dataset and are primarily included
for demonstration purposes. Indeed, manually writing enough components for
HELIX to support large-scale dataset generation is simply too much work for a
small team of researchers. Instead, we developed and implemented an approach
for automatically extracting HELIX components from existing open-source
libraries using program slicing.

Program slicing is an approach to divide a program into a minimal form that
still produces a desired behavior. It begins by selecting some desired behavior
and then removing code which does not influence that behavior based on code and
dataflow analysis \cite{weiser1981}. Program slicing has been used in a number
of different areas of program analysis, including debugging, program
comprehension, software maintenance, and software testing \cite{ngah2014}.

\begin{figure}
\begin{center}
\includegraphics[width=\linewidth]{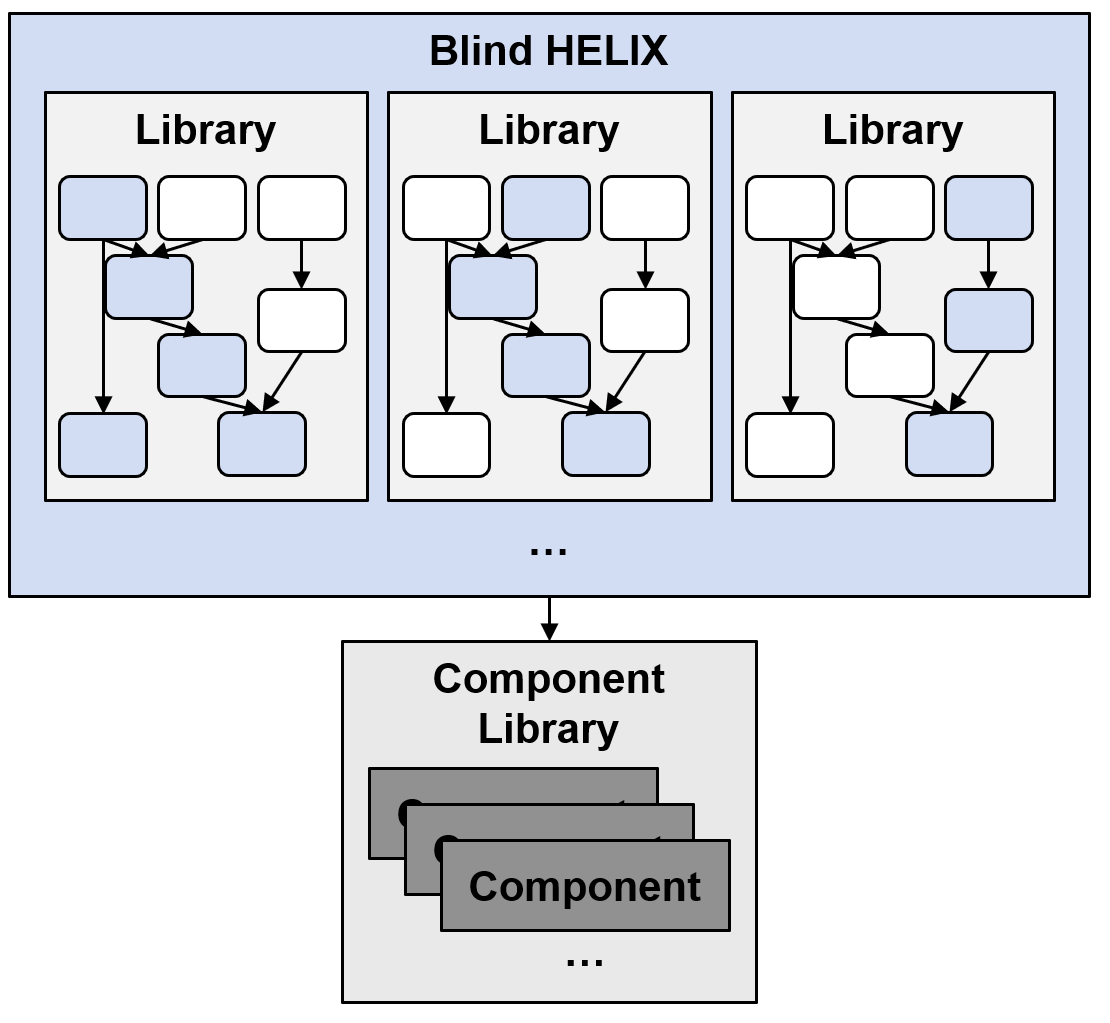}
\end{center}
\caption{\label{fig:blind-helix-program-slicing} Blind HELIX's use of program
slicing - a slice is taken starting at each exported function in the library to
generate a single HELIX component.}
\end{figure}

We implemented a program slicing approach to automatically extract HELIX
components from existing libraries, which we refer to as Blind HELIX. Blind
HELIX's design is depicted in Figure \ref{fig:blind-helix-program-slicing} and
it is built on top of the HELIX framework. Given a software library, Blind
HELIX uses the following procedure to extract HELIX components:

\begin{enumerate}
\item Parse the library metadata, identifying all exported functions.

\item For each exported function, attempt to build a binary which includes that
function and statically links against the target library. Link-time
optimization is enabled so that the linker handles slicing the library during a
dead code elimination pass.

\item Exported functions which fail to build are discarded.

\item A component is created for each successful build of an export. The built
binary is parsed and label tag added to the component for the name of each
included function.

\item The original library is packaged with all successfully identified and
labeled components and exported for later use. The library is rewritten to
rename all function symbols to avoid potential name collisions when linking
with multiple Blind HELIX components.
\end{enumerate}

Note that program slicing is handled automatically by the compiler in step 2.
By statically linking against a target library and only importing a single
function with link-time optimization enabled, a dead code elimination pass
removes unused functions and global data from the library. This leaves only the
minimal ``slice'' of the program responsible for that export's functionality.

Blind HELIX supports ELF libraries on Linux written in C but the approach is
generic enough to apply to Windows PEs and libraries written in C++.

\subsubsection*{A Corpus of Libraries}

We used VCPKG\footnote{VCPKG. \url{https://vcpkg.io/en/index.html}.} - a C and
C++ package manager for acquiring and managing libraries developed by
Microsoft. VCPKG supports libraries on Linux and Windows and is capable of
downloading libraries automatically and building them from source on any
platform it supports. As of the writing of this paper, VCPKG includes over
1,700 open-source libraries and Blind HELIX integrates with it directly to
download and build target libraries for component extraction.

After downloading and building every library supported by VCPKG, we used Blind
HELIX to extract components from compatible libraries written in C. We were
able to extract 28,178 unique components from 268 different libraries.

\subsubsection*{Labeling and Similarity}

In order to label individual components, we identified the set of functions
included in a given component's slice using the debugging symbols provided by
the library to which the component belongs. These function names make up a set
of component labels $L_c$ of strings in the form
\texttt{<library\_name>-<function\_name>} (e.g., \texttt{zlib-inflate}).

The set of labels for a particular sample $L_s$ built with $n$ components is
simply the union of the sets of lables in each component, as depicted in
Equation \ref{eq:sample-labels}.

\begin{equation}
L_s = \bigcup\limits_{i=1}^{n} L_{c,i}
\label{eq:sample-labels}
\end{equation}

To approximate the similarity between any two samples, we can take the Jaccard
similarity of their sets of labels as depicted in Equation \ref{eq:jaccard}.

\begin{equation}
S(L_{s1}, L_{s2}) = \frac{|L_{s1} \cap L_{s2}|}{|L_{s1} \cup L_{s2}|}
\label{eq:jaccard}
\end{equation}

\subsection{Dataset Generation}

There are a few potential pitfalls with naively combining extracted components
from the VCPKG libraries to generate datasets. First, some of the extracted
components are fundamentally incompatible with others and combinations
including both will fail to build. Despite library symbol renaming to avoid
link-time collisions, some inherent incompatibilities are unavoidable and
intrinsic to how the libraries themselves were written. To combat this, when
generating datasets from Blind HELIX components we simply attempt to generate
more than necessary and accept some loss due to build failures.

Second, randomly selecting included components from the corpus of available
components gives very little control over the distribution of ground truth
similarities in the generated dataset. This can be problematic if the dataset
generated is biased toward either extremely low or extremely high similarity,
since metrics which are also biased will perform unreasonably well when
evaluated against the dataset.

Finally, including more than one component from a single library in a single
build may lead to inflated levels of similarity not reflected in the sample
labels due to unintended function subgraph overlap captured when slicing the
library. To combat these issues, we implemented the following
library-stratified semi-random dataset generation strategy:

\begin{enumerate}
\item Randomly select a list of $n$ libraries, where $n$ is the number of
components to include in each sample.

\item Randomly select a slice component from each selected library.

\item Generate a sample from the list of selected components.

\item Randomly replace a random number of the selected components between zero
and $n * p$, where $p$ is a parameter that controls the over degree of
similarity in the generated dataset between zero and one. Components are
replaced in the same manner as 1 and 2, randomly selecting a replacement
library and then randomly selecting a slice component from that library.

\item Repeat steps 3 and 4, discarding samples which fail to build, until the
desired number of samples is reached.
\end{enumerate}

Using this strategy, and given that we were able to extract 28,178 unique
components from 268 different libraries, this approach can generate dataset
sizes which are, for all practical purposes, infinite\footnote{Assuming 50
components per sample, ${268\choose50} = 6.36 \times 10^{54}$ is a lower bound
on the possible number of samples that can be generated from VCPKG
components.}.

\section{Evaluation}
\label{sec:evaluation}

We evaluated Blind HELIX using a number of widely used program similarity tools
in the domain of malware analysis. We curated a dataset of programs
encompassing multiple, disparate notions of program similarity and manually
labeled the similarity between pairs of programs in the dataset. Finally, we
generated a synthetic dataset of 256 programs (32,640 distinct pairs) using
Blind HELIX and compared the relative ranking of program similarity tool
performance on the two datasets. In the next section, we describe each of the
program similarity tools in detail.

\subsection{Program Similarity Tools}

We selected five popular program similarity tools to use for this evaluation:
ssdeep, sdhash, the Trend Micro Locality-Sensitive Hash (TLSH), Lempel-Ziv
Jaccard Distance (LZJD) \cite{kornblum2006, roussev2010, oliver2013, raff2017},
and BinDiff\footnote{Zynamics BinDiff.
\url{https://www.zynamics.com/bindiff.html}.}.

ssdeep uses context-triggered piecewise hashing (CTPH) to generate
locality-sensitive fuzzy hashes for full files. Hashes that do not match
exactly are compared by computing a weighted edit distance between them to
produce a score between zero and one \cite{kornblum2006}. While originally
designed for assessing text similarity in an email spam filter algorithm,
ssdeep hashes are now commonly used for malware analysis and are featured on
VirusTotal\footnote{VirusTotal. \url{https://www.virustotal.com/}.}.

sdhash uses statically improbable byte sequences to generate hashes based on
empirical measurement of entropy in a dataset of various file types. Hashes are
compared by measuring their similarity as Bloom filters to produce a score
between zero and one \cite{roussev2010}.

TLSH uses a byte-wise sliding window approach to generate hashes, comparing
them by rough edit distance to produce an unbounded\footnote{For the purposes
of this analysis, we normalize TLSH scores between zero and one by selecting
optimal score bounds using the detection and false positive rates from the
original paper.} score where zero indicates that two files are identical and
higher values indicate lower similarity \cite{oliver2013}. 

LZJD uses the Lempel-Ziv technique for creating a compression dictionary of
previously-seen sub-sequences. They then use min-hashing as an optimization on
Jaccard distance between the sub-sequences of bytes for two files to compute a
similarity score between zero and one \cite{raff2017}.

BinDiff uses a large number of weighted similarity metrics based on structural
features of the code as well as features of the program's call graph aggregated
across the disassembly of the full program to generate a similarity score
between zero and one \cite{flake2004, dullien2005}. BinDiff differs from the
previous metrics in that it only works on executable files, not arbitrary blobs
of data, but it is the de facto standard for malware analysis and comparison.

We ran each of these program similarity metrics on every pair of programs in
the evaluation datasets and compared relative performance to the ground truth
similarity value for that pair. In the next section, we describe the datasets
these tools were evaluated against.

\subsection{Dataset}

In order to evaluate the quality of the datasets HELIX (and Blind HELIX) are
able to generate, we curated samples that represent the following abstract
notions of program similarity:

\begin{description}
\item[Versions] Different versions of the same source code. Samples are
binaries from GNU coreutils compiled from the source code versions released
many years apart.

\item[Optimizations] Different compilers and optimization levels applied to the
same source code. Samples are binaries from GNU coreutils and findutils
compiled with GCC and Clang at different optimization levels.

\item[Obfuscations] Different obfuscation techniques applied to the same source
code. Samples are example programs from the Android Native Development Kit
(NDK)\footnote{Android Native Development Kit.
\url{https://developer.android.com/ndk}.} compiled with various combinations of
obfuscations from the Obfuscator LLVM (OLLVM) compiler.

\item[Semantic] Different source code that accomplishes the same purpose.
Samples are selected from the Google Code Jam (GCJ)\footnote{Google Code Jam.
\url{https://codingcompetitions.withgoogle.com/codejam}.} programming
competition as solutions from different authors to the same challenge problem.

\item[Malware] Real-world malware from the same malware family. Samples include
different versions of components in the CosmicDuke malware family labeled by
compilation timestamp and originally identified by F-Secure \cite{fsecure2014}.
\end{description}

The selected categories represent real notions of program similarity toward
which many applications of program similarity are targeted. Pairs of samples in
each category were compared and given a score between zero and one by multiple
analysts familiar with program similarity and program analysis. Analysts were
given source code where applicable and disassembly otherwise. The scores given
by each analyst were then averaged to give the ground truth similarity labels
for the dataset. The full table of manually-labeled samples is included in
Appendix \ref{sec:manual-dataset-makeup} and the full dataset is included with
Blind HELIX at \url{https://github.com/helix-datasets/blind-helix}.

\begin{figure}
\begin{center}
\includegraphics[width=\linewidth]{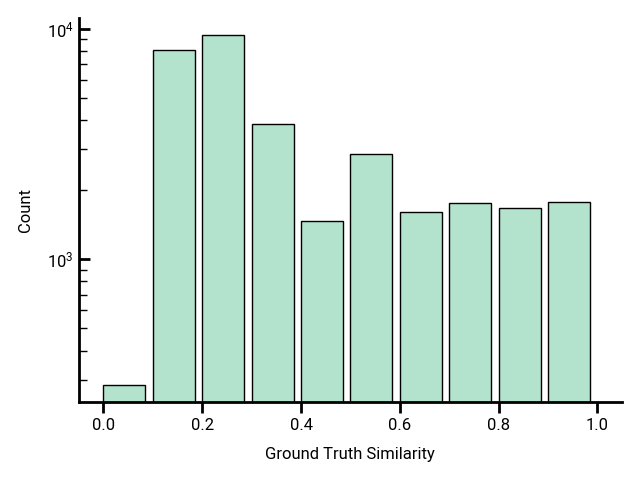}
\end{center}
\caption{\label{fig:helix-dataset-distribution} Distribution of ground truth
similarity scores in the HELIX-generated dataset. The dataset was generated
such that there would not be an unreasonable abundance of low-similarity or
high-similarity scores to avoid benefiting biased similarity metrics.}
\end{figure}

We then generated a separate dataset using HELIX for comparison, consisting of
256 samples for a total of 32,640 pairs. Samples were generated using
components automatically extracted from open source libraries in VCPKG using
Blind HELIX. The distribution of ground truth similarity values in the HELIX
dataset is depicted in Figure \ref{fig:helix-dataset-distribution}. Generating
this dataset took 2 minutes and 29 seconds running on an Intel Xeon 8253 CPU
with 64GB of RAM and dataset generation performance scales linearly with
dataset size.

In summary, we have two distinct datasets on which to evaluate program
similarity tools:

\begin{itemize}
\item A dataset of manually-labeled samples encompassing five different
abstract notions of program similarity.  Summarized in Appendix
\ref{sec:manual-dataset-makeup}.  Referred to as the ``manually-labeled
dataset'' for the remainder of this paper.

\item A dataset generated with HELIX from Blind HELIX components harvested from
VCPKG consisting of 256 samples (32,640 pairs). Referred to as the ``Blind
HELIX dataset'' or the ``HELIX dataset'' for the remainder of this paper.
\end{itemize}

\subsubsection*{Experiment}

Using our five program similarity tools we computed pairwise similarity scores
across both the manually-labeled and Blind HELIX datasets. We measured the
output of each of the tools against the ground truth similarity values provided
for each dataset by computing the mean absolute error (MAE) across all pairwise
comparisons. In the next section, we explain our results and explore some
specific examples of interesting performance by the similarity metrics.

\subsection{Results and Analysis}

\begin{figure}
\begin{center}
\includegraphics[width=\linewidth]{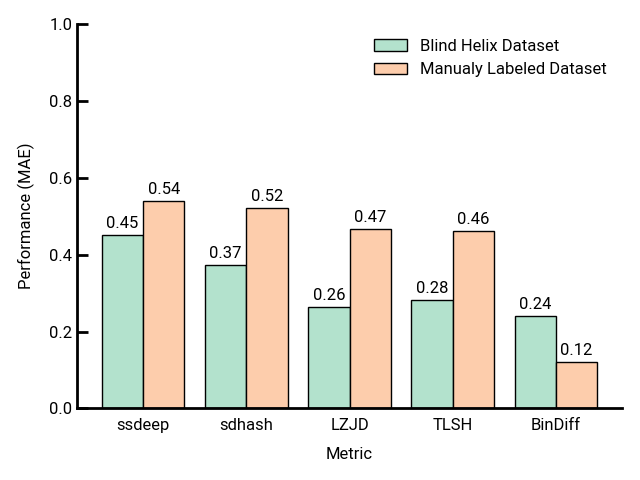}
\end{center}
\caption{\label{fig:performance-manual-vs-helix} Program similarity metric
performance on the Blind HELIX and manually-labeled datasets - lower is
better.}
\end{figure}

The performance of each of the five program similarity tools is depicted in
Figure \ref{fig:performance-manual-vs-helix}. On the manually-labeled dataset
the worst performing metric is ssdeep very closely followed by sdhash. LZJD and
TLSH perform a bit better and fairly comparably to one another. Finally,
BinDiff performs the best by a fairly large margin. The relative program
similarity tool rankings are essentially the same on the Blind HELIX dataset
with ssdeep and sdhash performing the worst, followed by LZJD and TLSH which
perform fairly comparably to one another, and finally BinDiff again performs
the best. The only difference between the relative ranking of the tools
compared across both datasets is the position of LZJD and TLSH which are very
close in performance - within 0.02 MAE of each other on both datasets.

From this similar performance ranking of program similarity tools we postulate
that Blind HELIX datasets embody the same notions of program similarity
represented by the manually-labeled dataset. In other words, datasets generated
by Blind HELIX correctly model realistic and useful notions of program
similarity and can be a useful tool for evaluating the effectiveness of new
program similarity approaches.

\subsubsection*{Manually-Labeled Dataset Performance}

\begin{figure*}[t]
\begin{center}
\includegraphics[width=\linewidth]{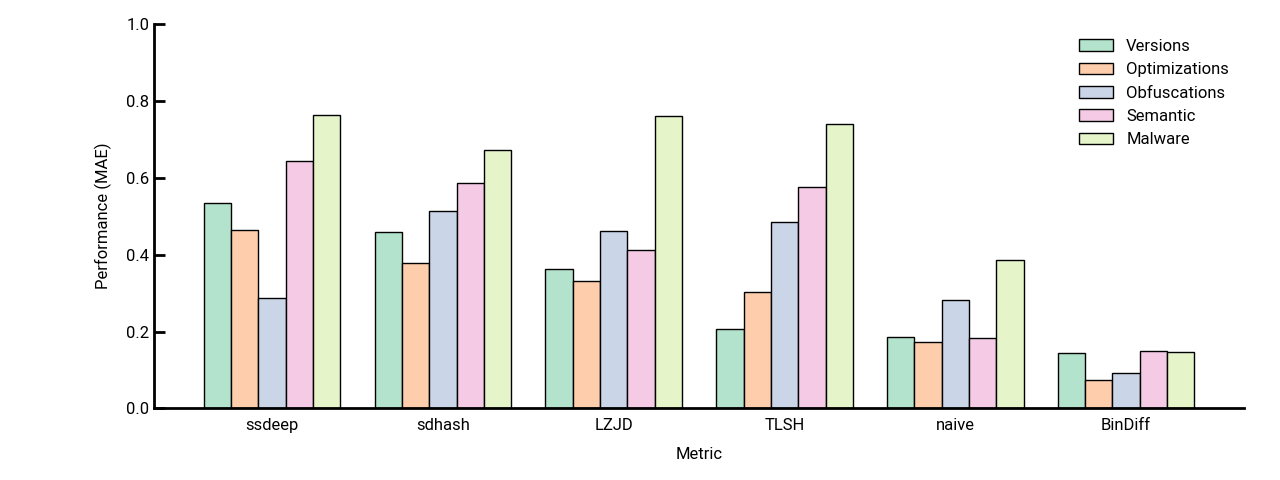}
\end{center}
\caption{\label{fig:performance-manual-comparative} Comparative program
similarity metric performance on manually-labeled dataset similarity categories
- lower is better. The ``naive'' metric represents always guessing a similarity
of 0.5.}
\end{figure*}

The comparative performance of each of the five program similarity tools on
individual categories of program similarity within the manually-labeled dataset
is depicted in Figure \ref{fig:performance-manual-comparative}. There are a few
interesting insights we can draw about each of the tools from the results on
this dataset.

First, while LZJD and TLSH perform relatively comparably in the context of the
entire dataset, LZJD appears to be better suited to identifying semantic
similarity while TLSH seems better suited to identifying similarities across
different versions of software. This could be because TLSH's quartile
computation leads it to perform well against data that varies significantly in
only a few places. Differing versions of large programs are likely to only have
modifications or additions to a few functions resulting in localized changes,
while semantically similar programs built from differing source code are likely
to have a diffuse, low level of dissimilarity throughout.

Second, BinDiff appears to perform somewhat better on optimizations and
obfuscations than other categories. This could be due to its focus on call
graph comparison and aggregation of basic block level similarity, thus it is
optimized to find similarities in call graph structure and basic units of
functionality. These would likely be preserved by a compiler or an obfuscator
which must necessarily preserve program functionality. In contrast, it performs
the worst on semantic similarity where individual basic blocks, call graph
structure, and data flow may vary significantly between samples while
functionality is preserved.

Finally, it's worth noting that metrics like ssdeep and sdhash are primarily
designed to discover similarities between nearly identical files and were
originally built for comparing textual data, not complex binary formats
consisting of a mix of metadata and machine code with no specific alignment.
Indeed, across the entire manually-labeled dataset, ssdeep reports a similarity
value of zero for 93.4\% of pairs.  As a point of comparison, Figure
\ref{fig:performance-manual-comparative} includes a ``naive'' metric that
simply always predicts a similarity value of 0.5. This metric is obviously
useless and outperforms nearly all of the program similarity tools yet these
tools remain widely used for program similarity.

\subsubsection*{Case Study: Malware Samples}

\begin{figure}
\begin{center}
\includegraphics[width=\linewidth]{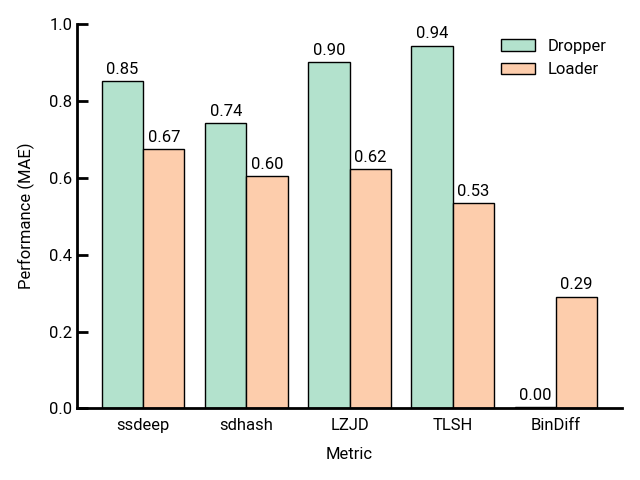}
\end{center}
\caption{\label{fig:performance-malware} Program similarity metric performance
on CosmicDuke malware samples by sample category - lower is better.}
\end{figure}

With the exception of BinDiff, all metrics exhibited the weakest performance on
the malware category of similarity. This portion of the dataset consisted of
malware samples from the family CosmicDuke as reported by F-Secure. These
samples catalog the development of a dropper and a loader component of the
CosmicDuke malware family over time and are differentiated by their compilation
timestamps in the original report. For samples belonging to the dropper
component of CosmicDuke, the executable code is actually identical and the
compilation timestamps match; the only difference between these samples is
binary metadata designed to trick the target into running the malware.

Manual analysts were able to easily identify the identical programs and gave
them a very high ground truth similarity in our manually-labeled dataset but,
as depicted in Figure \ref{fig:performance-malware}, BinDiff was the only
metric that was able to identify and properly weight the importance of this
similarity. This is likely because the other metrics operate on raw bytes,
native of the binary executable format and so overweighted the importance of
the program metadata in identifying similarities between samples.

\section{Discussion and Future Work}
\label{sec:discussion}

In this section, we discuss the results of our evaluation, some limitations,
and describe some directions for future work.

\subsubsection*{Limitations}

As of the writing of this paper, HELIX only includes a relatively small number
of hand-written components. Blind HELIX is one solution to creating a large
number of HELIX components for dataset generation but has a few limitations.
First, Blind HELIX currently only supports libraries written in C for Linux,
but could feasibly be extended to support C++ and Windows. Second, Blind HELIX
labels are not based on program functionality, just a rough approximation of it
through included functions in the library slice subgraph. For the same reason,
when Blind HELIX components call their exported library function they are not
passed reasonable or realistic function arguments. If executed, samples built
with Blind HELIX would mostly likely crash immediately. HELIX is actively
maintained and we invite community input and contribution of additional
components and tools built on top of HELIX that improve upon Blind HELIX's
automated component extraction.

\subsubsection*{Classification Datasets}

In this work, we focused on generating samples for program similarity using
HELIX and Blind HELIX. However, the HELIX framework is flexible enough to be
used for more complex classification dataset generation to aid in research
applying machine learning to program analysis. Given that, with Blind HELIX, we
can generate two programs with arbitrary levels of similarity, it should be
possible to generate synthetic ``classes'' of programs with configurable
statistical distribution parameters to suit the needs of any particular
evaluation. We leave this to future work.

\subsubsection*{Dynamic Analysis}

As mentioned before, samples generated from Blind HELIX components are not
guaranteed to function when executed and would mostly likely crash. This makes
Blind HELIX datasets somewhat useless for applications of dynamic analysis,
however this limitation does not apply to the core HELIX framework. HELIX
itself includes a number of components that accomplish interesting and useful
functions all labeled according to the MITRE ATT\&CK framework\footnote{MITRE
ATT\&CK.  \url{https://attack.mitre.org/}.}. Unfortunately these hand-written
components take time and resources to develop and so only a relatively small
number exist as of the writing of this paper. Hopefully a community effort to
contribute functional, well-labeled components will improve HELIX's dynamic
analysis dataset generation capabilities in the future.

\section{Related Work}
\label{sec:related-work}

There is some related work in the area of automated evaluation of similarity
techniques. Breitinger et al. built a framework to test algorithms of
similarity matching (or FRASH, for short) which takes as input a test file or
files and mutates those files as input data to evaluate the performance of the
ssdeep and sdhash tools \cite{breitinger2013}. HELIX allows users to create raw
datasets which can be used for program similarity research and is not
restricted to evaluating only ssdeep and sdhash. Additionally, the mutations
performed by FRASH do not guarantee that test cases will still be functional
executables. Breitinger et al. improve on FRASH in \cite{breitinger2014},
automatically approximating similarity of real-world data by using longest
common subsequence approximation. Their evaluation was conducted on
non-executable file formats (images, html, PDFs, etc.), however, and it's not
clear that this approach would directly apply to much more complex binary
executable file formats.

In the domain of malware analysis, multiple similarity assessment frameworks
have been proposed. Upchurch et al. proposed a framework for malware similarity
testing entitled Variant \cite{upchurch2015}. Variant, however, consists of a
single static dataset composed of only 85 samples of malware manually analyzed
and grouped by subject matter experts. Although HELIX's samples are synthetic,
not real-world malware, HELIX is capable of generating much larger datasets
with high-quality ground truth labels based on real program similarity. Liang
et al.  extracted call traces from malware executions and proposed a method of
winnowing a large dataset into one which could be used as a benchmark. Their
approach is purely focused on dynamic analysis and labels samples based on a
``maliciousness index'' computed via a voting system of antivirus labels.
Sebasti\'an et al. proposed a method of automatically labeling real-world
malware by resolving discrepancies between the labels provided by different
antivirus vendors for the same sample \cite{sebastian2016}. However, unlike
HELIX, antivirus labels are largely opaque and rarely connected to any real
notion of semantic, structural, or behavioral similarity.

Toward automated program variant generation, Choi et al. proposed a system for
applying non-behavior-modifying transformations to source code repeatedly,
guided by a genetic algorithm until a specified level of diversity is achieved.
This work takes the opposite approach to ours and produces many diverse samples
from a single piece of source code \cite{choi2019}. Their adaptive malware
variant generation (AMVG) framework is presented as a proof of concept and not
designed for dataset generation, though it may be interesting to apply some of
their genetic algorithm approach to HELIX in future work.

\section{Conclusion}
\label{sec:conclusion}

Dataset quality and general availability is a major problem in the domain of
program similarity research. In this work, we have demonstrated a practical
approach for generating synthetic program similarity datasets from existing
library code using program slicing and recombination that is capable of
generating practically infinitely sized datasets. In our evaluation, we have
shown that our approach correctly models realistic and useful notions of
program similarity. Finally, we have open-sourced our framework for program
generation and mutation (HELIX) and our tool for automatic extraction of
functional components from existing libraries (Blind HELIX).

\subsection*{Acknowledgements}

We would like to thank all of our colleagues at MIT Lincoln Laboratory that
have contributed to the conception and development of this work along the way.

This material is based upon work supported by the Department of Defense under
Air Force Contract No. FA8721-05-C-0002 and/or FA8702-15-D-0001. Any opinions,
findings, conclusions or recommendations expressed in this material are those
of the author(s) and do not necessarily reflect the views of the Department of
Defense.

\subsection*{Availability}

HELIX (the program generation/mutation framework) was developed for this
research and is publicly available at
\url{https://github.com/helix-datasets/helix}. Blind HELIX (the tool to
automatically extract HELIX components from library code using program slicing)
is also publicly available at
\url{https://github.com/helix-datasets/blind-helix}. The small,
manually-labeled evaluation dataset, as well as the source code for all scripts
used for evaluation are available in the Blind HELIX
repository\footnote{Malware samples are included by hash only.}.

\bibliographystyle{plain}
\bibliography{references}

\begin{thebibliography}{10}

\bibitem{fsecure2014}
{CosmicDuke}: {Cosmu} with a twist of {MiniDuke}.
\newblock Technical report, {F-Secure Labs}, 2014.

\bibitem{anderson2018}
Hyrum~S. Anderson and Phil Roth.
\newblock Ember: An open dataset for training static pe malware machine
  learning models.
\newblock {\em arXiv e-prints}, 2018.

\bibitem{arp2014}
Daniel Arp, Michael Spreitzenbarth, Malte Hubner, Hugo Gascon, Konrad Rieck,
  and CERT Siemens.
\newblock Drebin: Effective and explainable detection of android malware in
  your pocket.
\newblock In {\em Ndss}, volume~14, pages 23--26, 2014.

\bibitem{avino2018}
Laura Avi{\~n}{\'o}, Matteo Ruffini, and Ricard Gavald{\`a}.
\newblock Generating synthetic but plausible healthcare record datasets.
\newblock {\em arXiv preprint arXiv:1807.01514}, 2018.

\bibitem{benaim2020}
Anat~Reiner Benaim, Ronit Almog, Yuri Gorelik, Irit Hochberg, Laila Nassar,
  Tanya Mashiach, Mogher Khamaisi, Yael Lurie, Zaher~S Azzam, Johad Khoury,
  et~al.
\newblock Analyzing medical research results based on synthetic data and their
  relation to real data results: systematic comparison from five observational
  studies.
\newblock {\em JMIR medical informatics}, 8(2):e16492, 2020.

\bibitem{botacin2021}
Marcus Botacin, Fabricio Ceschin, Ruimin Sun, Daniela Oliveira, and Andr{\'e}
  Gr{\'e}gio.
\newblock Challenges and pitfalls in malware research.
\newblock {\em Computers \& Security}, 106:102287, 2021.

\bibitem{breitinger2014}
Frank Breitinger and Vassil Roussev.
\newblock Automated evaluation of approximate matching algorithms on real data.
\newblock In {\em Digital Investigation}, volume~11, pages S10--S17, 2014.

\bibitem{breitinger2013}
Frank Breitinger, Georgios Stivaktakis, and Harald Baier.
\newblock Frash: A framework to test algorithms of similarity hashing.
\newblock In {\em Digital Investigation}, volume~10, pages S50--S58, 2013.

\bibitem{choi2019}
Jusop Choi, Dongsoon Shin, Hyoungshick Kim, Jason Seotis, and Jin~B Hong.
\newblock Amvg: Adaptive malware variant generation framework using machine
  learning.
\newblock In {\em 2019 IEEE 24th Pacific Rim International Symposium on
  Dependable Computing (PRDC)}, pages 246--24609, 2019.

\bibitem{dash2019}
Saloni Dash, Ritik Dutta, Isabelle Guyon, Adrien Pavao, Andrew Yale, and
  Kristin~P Bennett.
\newblock Synthetic event time series health data generation.
\newblock {\em arXiv preprint arXiv:1911.06411}, 2019.

\bibitem{dolan2016}
Brendan Dolan-Gavitt, Patrick Hulin, Engin Kirda, Tim Leek, Andrea Mambretti,
  Wil Robertson, Frederick Ulrich, and Ryan Whelan.
\newblock Lava: Large-scale automated vulnerability addition.
\newblock In {\em 2016 IEEE Symposium on Security and Privacy (SP)}, pages
  110--121. IEEE, 2016.

\bibitem{dullien2005}
Thomas Dullien and Rolf Rolles.
\newblock Graph-based comparison of executable objects.
\newblock 2005.

\bibitem{flake2004}
Halvar Flake.
\newblock Structural comparison of executable objects.
\newblock 2004.

\bibitem{grajeda2017}
Cinthya Grajeda, Frank Breitinger, and Ibrahim Baggili.
\newblock Availability of datasets for digital forensics--and what is missing.
\newblock {\em Digital Investigation}, 22:S94--S105, 2017.

\bibitem{gupta2016}
Ankush Gupta, Andrea Vedaldi, and Andrew Zisserman.
\newblock Synthetic data for text localisation in natural images.
\newblock In {\em Proceedings of the IEEE conference on computer vision and
  pattern recognition}, pages 2315--2324, 2016.

\bibitem{haq2019}
Irfan~Ul Haq and Juan Caballero.
\newblock A survey of binary code similarity.
\newblock {\em arXiv preprint arXiv:1909.11424}, 2019.

\bibitem{harichandran2016}
Vikram~S Harichandran, Frank Breitinger, and Ibrahim Baggili.
\newblock Bytewise approximate matching: the good, the bad, and the unknown.
\newblock {\em Journal of Digital Forensics, Security and Law}, 11(2):4, 2016.

\bibitem{hittmeir2019}
Markus Hittmeir, Andreas Ekelhart, and Rudolf Mayer.
\newblock On the utility of synthetic data: An empirical evaluation on machine
  learning tasks.
\newblock In {\em Proceedings of the 14th International Conference on
  Availability, Reliability and Security}, pages 1--6, 2019.

\bibitem{jaderberg2014}
Max Jaderberg, Karen Simonyan, Andrea Vedaldi, and Andrew Zisserman.
\newblock Synthetic data and artificial neural networks for natural scene text
  recognition.
\newblock {\em arXiv preprint arXiv:1406.2227}, 2014.

\bibitem{kornblum2006}
Jesse Kornblum.
\newblock Identifying almost identical files using context triggered piecewise
  hashing.
\newblock In {\em Proceedings of the 6th Annual Digital Forensic Research
  Workshop (DFRWS '06)}, pages 91--97, 2006.

\bibitem{nappa2013}
Antonio Nappa, M~Zubair Rafique, and Juan Caballero.
\newblock Driving in the cloud: An analysis of drive-by download operations and
  abuse reporting.
\newblock In {\em International Conference on Detection of Intrusions and
  Malware, and Vulnerability Assessment}, pages 1--20. Springer, 2013.

\bibitem{ngah2014}
Amir Ngah and Siti~Aminah Selamat.
\newblock A brief survey of program slicing.
\newblock In {\em International Symposium on Research in Innovation and
  Sustainability 2014 (ISoRIS’14)}, pages 1467--1470. Citeseer, 2014.

\bibitem{nikolenko2019}
Sergey~I Nikolenko et~al.
\newblock Synthetic data for deep learning.
\newblock {\em arXiv preprint arXiv:1909.11512}, 3, 2019.

\bibitem{oliver2013}
Jonathan Oliver, Chun Cheng, and Yanggui Chen.
\newblock {TLSH} - a locality sensitive hash.
\newblock {\em 4th Cybercrime and Trustworthy Computing Workshop}, 2013.

\bibitem{pagani2018}
Fabio Pagani, Matteo Dell'Amico, and Davide Balzarotti.
\newblock Beyond precision and recall: understanding uses (and misuses) of
  similarity hashes in binary analysis.
\newblock In {\em Proceedings of the Eighth ACM Conference on Data and
  Application Security and Privacy}, pages 354--365, 2018.

\bibitem{patki2016}
Neha Patki, Roy Wedge, and Kalyan Veeramachaneni.
\newblock The synthetic data vault.
\newblock In {\em 2016 IEEE International Conference on Data Science and
  Advanced Analytics (DSAA)}, pages 399--410. IEEE, 2016.

\bibitem{raff2017}
Edward Raff and Charles Nicolas.
\newblock An alternative to ncd for large sequences, lempel-ziv jaccard
  distance.
\newblock In {\em Proceedings of the 23rd ACM SIGKDD International Conference
  on Knowledge Discovery and Data Mining}, pages 1007--1015, 2017.

\bibitem{ronan2018}
Royi Ronan, Marian Radu, Corina Feuerstein, Elad Yom-Tov, and Mansour Ahmadi.
\newblock Microsoft malware classification challenge.
\newblock {\em arXiv e-prints}, 2018.

\bibitem{roussev2010}
Vassil Roussev.
\newblock Data fingerprinting with similarity digests.
\newblock In K.~Chow and S.~Shenoi, editors, {\em Research Advances in Digital
  Forensics VI}, pages 207--226. Springer, 2010.

\bibitem{roussev2011}
Vassil Roussev.
\newblock An evaluation of forensic similarity hashes.
\newblock {\em digital investigation}, 8:S34--S41, 2011.

\bibitem{roussev2013}
Vassil Roussev and Candice Quates.
\newblock File fragment encoding classification—an empirical approach.
\newblock {\em Digital Investigation}, 10:S69--S77, 2013.

\bibitem{sebastian2016}
Marcos Sebasti{\'a}n, Richard Rivera, Platon Kotzias, and Juan Caballero.
\newblock {AVClass}: A tool for massive malware labeling.
\newblock In {\em International symposium on research in attacks, intrusions,
  and defenses}, pages 230--253. Springer, 2016.

\bibitem{skopik2014}
Florian Skopik, Giuseppe Settanni, Roman Fiedler, and Ivo Friedberg.
\newblock Semi-synthetic data set generation for security software evaluation.
\newblock In {\em 2014 Twelfth Annual International Conference on Privacy,
  Security and Trust}, pages 156--163. IEEE, 2014.

\bibitem{smith2020}
Michael~R. Smith, Nicolas~T. Johnson, Joe~B. Ingram, Armida~J. Carbajal,
  Bridget~I. Haus, Eva Domscot, Ramyaa Ramyaa, Christopher~C. Lamb, Stephen~J.
  Verzi, and W.~Philip Kegelmeyer.
\newblock Mind the gap: On bridging the semantic gap between machine learning
  and malware analysis.
\newblock In {\em Proceedings of the 13th ACM Workshop on Artificial
  Intelligence and Security (AISec 2020)}, pages 49--60, 2020.

\bibitem{tremblay2018}
Jonathan Tremblay, Aayush Prakash, David Acuna, Mark Brophy, Varun Jampani, Cem
  Anil, Thang To, Eric Cameracci, Shaad Boochoon, and Stan Birchfield.
\newblock Training deep networks with synthetic data: Bridging the reality gap
  by domain randomization.
\newblock In {\em Proceedings of the IEEE conference on computer vision and
  pattern recognition workshops}, pages 969--977, 2018.

\bibitem{ucci2019}
Daniele Ucci, Leonardo Aniello, and Roberto Baldoni.
\newblock Survey of machine learning techniques for malware analysis.
\newblock {\em Computers \& Security}, 81:123--147, 2019.

\bibitem{upchurch2015}
Jason Upchurch and Xiaobo Zhou.
\newblock Variant: a malware similarity testing framework.
\newblock In {\em 2015 10th International Conference on Malicious and Unwanted
  Software (MALWARE)}, pages 31--39, 2015.

\bibitem{weiser1981}
Mark Weiser.
\newblock Program slicing.
\newblock In {\em Proceedings of the 5th international conference on Software
  Engineering (ICSE '81)}, pages 43--53, 1981.

\bibitem{yale2019}
Andrew Yale, Saloni Dash, Ritik Dutta, Isabelle Guyon, Adrien Pavao, and
  Kristin Bennett.
\newblock Privacy preserving synthetic health data.
\newblock In {\em ESANN 2019-European Symposium on Artificial Neural Networks,
  Computational Intelligence and Machine Learning}, 2019.

\bibitem{zheng2018}
Muwei Zheng, Hannah Robbins, Zimo Chai, Prakash Thapa, and Tyler Moore.
\newblock Cybersecurity research datasets: taxonomy and empirical analysis.
\newblock In {\em 11th {USENIX} Workshop on Cyber Security Experimentation and
  Test ({CSET} 18)}, 2018.

\bibitem{zhou2012}
Yajin Zhou and Xuxian Jiang.
\newblock Dissecting android malware: Characterization and evolution.
\newblock In {\em 2012 IEEE symposium on security and privacy}, pages 95--109.
  IEEE, 2012.

\end{thebibliography}

\appendix

\section{Manually-Labeled Dataset Makeup}
\label{sec:manual-dataset-makeup}

The manually labeled dataset samples are included in Tables
\ref{tab:manual-dataset-makeup-versions},
\ref{tab:manual-dataset-makeup-optimizations},
\ref{tab:manual-dataset-makeup-obfuscations},
\ref{tab:manual-dataset-makeup-semantic}, and
\ref{tab:manual-dataset-makeup-malware}.

\begin{table*}[t]
\begin{center}
\begin{tabular}{ c c c c }
\hline
Category & Class(s) & Description & Sample(s) \\
\hline
Versions & base64, du, ls, mv & GNU Coreutils version 6.12 released in 2008 & v6.12 \\
& & GNU Coreutils version 7.6 released in 2009 & v7.6 \\
& & GNU Coreutils version 8.32 released in 2020 & v8.32 \\
& & GNU Coreutils version 9.0 released in 2021 & v9.0 \\
\hline
\end{tabular}
\end{center}
\caption{\label{tab:manual-dataset-makeup-versions} Manually-labeled samples of
different versions of the same program.}
\end{table*}

\begin{table*}[t]
\begin{center}
\begin{tabular}{ c c c c }
\hline
Category & Class(s) & Description & Sample(s) \\
\hline
Optimizations & base64, du, ls, find & GNU Coreutils compiled with GCC at O0 & GCC O0\\
& & GNU Coreutils compiled with GCC at O1 & GCC O1\\
& & GNU Coreutils compiled with GCC at O2 & GCC O2\\
& & GNU Coreutils compiled with GCC at O3 & GCC O3\\
& & GNU Coreutils compiled with GCC at Of & GCC Of\\
& & GNU Coreutils compiled with GCC at Os & GCC Os\\
& & GNU Coreutils compiled with Clang at O0 & Clang O0\\
& & GNU Coreutils compiled with Clang at O2 & Clang O2\\
& & GNU Coreutils compiled with Clang at Of & Clang Of\\
\hline
\end{tabular}
\end{center}
\caption{\label{tab:manual-dataset-makeup-optimizations} Manually-labeled
samples of the same code compiled with different optimization levels and
compilers.}
\end{table*}

\begin{table*}[t]
\begin{center}
\begin{tabular}{ c c c c }
\hline
Category & Class(s) & Description & Sample(s) \\
\hline
Obfuscations & ollvm-ndk-native-plasma & Renders a plasma effect in a bitmap in C & unobf, fla, bcf, fla+bcf+sub \\
& ollvm-ndk-camera-texture-view & Preview NDK camera image & unobf, fla \\
& ollvm-ndk-endless-tunnel & A simple game implemented in native code & unobf, sub, fla, sub+fla \\
& ollvm-ndk-webp-view & Decode and rotate WebP images & unobf, sub, fla, sub+fla \\
\hline
\end{tabular}
\end{center}
\caption{\label{tab:manual-dataset-makeup-obfuscations} Manually-labeled
samples of different obfuscations of the same source code.}
\end{table*}

\begin{table*}[t]
\begin{center}
\begin{tabular}{ c c c c }
\hline
Category & Class(s) & Description & Sample(s) \\
\hline
Semantic & theme-park & GCJ 2010 Qualification & Ferlon, johny500, Prahadeesh, Suyog, Tommalla \\
& perfect-game & GCJ 2012 Round 3 & Aleksei, iPeter, Xhark, yangzhe1991, ytau \\
& nile & GCJ 2014 Round 2 & bwps, daimi89, LeeSin, Mosa \\
& recalculating & "GCJ 2020 Round 3 & betrue12, Bodo171, MicGor, minQZ, walnutwaldo20 \\
\hline
\end{tabular}
\end{center}
\caption{\label{tab:manual-dataset-makeup-semantic} Manually-labeled samples of
different semantically equivalent programs with different source code.}
\end{table*}

\begin{table*}[t]
\begin{center}
\begin{tabular}{ c c c c }
\hline
Category & Class(s) & Description & Sample(s) \\
\hline
Malware & cosmicduke-loader & CosmicDuke loader & 20120727-764add69922342b8c4200d64652fbee1376adf1c \\
& & & 20121113-4e3c9d7eb8302739e6931a3b5b605efe8f211e51 \\
& & & 20121113-580eca9e36dcd1a2deb9075bcae90afee46aace2 \\
& & & 20121113-5c5ec0b5112a74a95edc23ef093792eb3698320e \\
& & & 20121113-ccb29875222527af4e58b9dd8994c3c7ef617fd8 \\
& & & 20121204-9700c8a41a929449cfba6567a648e9c5e4a14e70 \\
& & & 20140418-fecdba1d903a51499a3953b4df1d850fbd5438bd \\
Malware & cosmicduke-dropper & CosmicDuke dropper & 0e5f55676e01d8e41d77cdc43489da8381b68086 \\
& & & 5a199a75411047903b7ba7851bf705ec545f6da9 \\
& & & 7631f1db92e61504596790057ce674ee90570755 \\
& & & f621ec1b363e13dd60474fcfab374b8570ede4de \\
\hline
\end{tabular}
\end{center}
\caption{\label{tab:manual-dataset-makeup-malware} Manually-labeled samples of
malware from the same family.}
\end{table*}

\end{document}